# Complex pattern in periodically driven binary Quincke colloids

Chuan Yu Zhang,* Mu Yue Zhao,* Xiao Yi Zhou,* Wen-De Tian,† Kang Chen,‡ and Tian Hui Zhang§
*Center for Soft Condensed Matter Physics and Interdisciplinary Research & School of Physical Science and Technology, Soochow University, Suzhou 215006, China*

Binary active systems offer a minimal platform for understanding how component heterogeneity organizes non-equilibrium structures. Here we show that in a binary Quincke system driven by a periodic square-wave electric field, a dynamic square lattice in which small particles occupy the lattice-cell centers which are surrounded by large particles emerges spontaneously. Particle-resolved trajectories reveal that the size separation arises from the mobility difference of large particles and small particles. Increasing the driving frequency reduces the single-cycle path lengths of both species and simultaneously decreases the lattice constant. These results show how size-dependent activity, density redistribution, and periodic forcing cooperate to generate tunable complex active patterns.

*Introduction.*— Self-assembly provides a fundamental route for constructing functional structures from molecular, colloidal, and nanoscale building blocks through local interactions, while directed and symmetry-breaking assembly further expand the accessible material states[1–3]. Many structures in living and driven systems, however, are maintained by continuous energy consumption, dissipation, and rearrangement rather than by relaxation to equilibrium[4–6]. Active matter is a representative class of such nonequilibrium systems, in which energy-consuming constituents generate persistent motion and, through interactions and fluctuations, display collective transport, phase separation, dynamic clustering, and pattern formation[4–10]. Artificial active colloids are especially useful because their size, propulsion, interactions, confinement, and external driving can be controlled independently, enabling particle-resolved studies of nonequilibrium many-body organization[7, 8, 11, 12].

A central question is how heterogeneity among coexisting components modifies active self-organization. Biological functional structures commonly arise from coupled interactions among different constituents; for example, actin and vimentin form interpenetrating networks that jointly regulate cellular mechanics[13]. Multicomponent active-colloid systems provide minimal and controllable models for isolating such effects[14–18]. In binary active populations with different particle sizes, motility differences and nonreciprocal interactions can produce partial demixing and self-sorting[19], while alternating driving can enhance directional memory and generate propagating activity waves[20]. More broadly, motility-induced phase separation and experiments on self-propelled colloids show that changes in particle mobility can strongly couple to density redistribution and phase organization[15, 21–23]. How size heterogeneity cooperates with time-dependent driving to organize ordered multicomponent structures, however, remains insufficiently understood.

Electrically driven Quincke rollers provide a suitable platform for addressing this problem. Above the Quincke threshold, unstable interfacial polarization drives spontaneous rotation of dielectric particles, which is converted near an electrode into translational rolling motion. Roller populations can develop polar motion and confined vortices through activity, alignment, repulsion, hydrodynamic and electrostatic interactions, and boundary effects[24–28]. Particle shape further tunes chiral motion and collective dynamics in ellipsoidal rollers[29], including flipping-assisted alignment[30] and large-scale vortices under high-frequency driving[31]. Attractive interactions can also promote polar clustering[32]. Their activity can also be rapidly controlled by the electric field. In micellar fluids, square-wave driving also produces frequency-dependent memory and activity[33]. Previous studies have demonstrated reconfigurable chiral patterns and collective excitable structures[34], ordered lattice assembly under pulsed or periodically modulated fields[35], shock waves[36], and activity waves in condensed active liquids and active crystals[37]. Periodic density feedback has also been reported to generate self-dual stripe and square-lattice patterns[38]. Most temporally programmed studies, however, have focused on monodisperse populations.

Here we investigate a binary suspension of 6 $\mu$m and 10 $\mu$m Quincke rollers under periodic square-wave electric-field driving. We show that periodic forcing couples size-dependent particle dynamics to finite-scale spatial demixing and a dynamic square lattice, with small particles occupying lattice-cell centers and large particles surrounding the periphery. We further quantify the fourfold structural order and show that increasing the driving frequency reduces the single-cycle path lengths of both species and decreases the lattice constant, demonstrating tunable pattern formation through temporal driving.

*Results.*—

The experiments use a binary suspension of 6 $\mu$m and 10 $\mu$m polystyrene particles with a number ratio of 1:1. A periodic square-wave electric field switches between a peak field $E_p > E_c$ and zero, with equal on and off durations of T/2. Unless otherwise specified, the representative data are obtained at $E_p/E_c = 2$, f = 20 Hz, and global area fraction $\varphi = 0.175$. At 20 Hz the field-off interval is 25 ms, much longer than the characteristic velocity-relaxation time (about 4 ms) and polarization-charge relaxation time (about 2 ms). Thus the particle velocity and polarization relax during every field-off stage, whereas the instantaneous spatial density distribution is retained for the next activation stage.

Exploratory measurements at different particle concentrations indicate that the dynamic square-lattice state is favored within an intermediate-density regime. At relatively high area fractions, particle redistribution becomes increasingly hindered and clear compositional demixing is difficult to establish, whereas at sufficiently low area fractions the particle density is insufficient to form a well-defined lattice network. The quantitative analysis in the present work therefore focuses on the representative intermediate-density condition $\varphi = 0.175$. The 6 $\mu$m:10 $\mu$m number ratio was fixed at 1:1 and was not systematically varied in the present study.

Figure 1 and Movie 1 show the evolution from the initially mixed suspension to the spatially organized state. At 0T, the two species are approximately randomly and uniformly mixed [Fig. 1(a)]. Repeated electric-field activation generates local density fluctuations and irregular high- and low-density regions [Fig. 1(b)]. During the following cycles, particles activated inside dense regions experience strong mutual electrostatic repulsion[27] and redistribute toward neighboring dilute regions. This repeated density redistribution progressively amplifies the compositional contrast. By 182T, a pronounced periodic pattern has emerged [Fig. 1(c)]: the 6 $\mu$m particles preferentially occupy the dense lattice-cell centers, whereas the 10 $\mu$m particles are enriched around the cell peripheries.

The compositional evolution is quantified by the scale-dependent segregation parameter D(L) [Fig. 1(d)]. Relative to the initial state, the curves shift upward at 50T and 182T, and the crossing with the operational threshold D(L) = 0.5 moves toward larger L. Thus both the magnitude and the characteristic length of the local segregation increase during pattern formation. At sufficiently large coarse-graining scales, D(L) decreases because each large observation box contains both particle species. The segregation is therefore finite in scale rather than a macroscopic complete separation of the binary components. A sensitivity analysis using nearby thresholds D = 0.4, 0.5, and 0.6 gives the same increasing trend of the characteristic segregation scale (Table S1), showing that this conclusion is insensitive to the precise operational threshold.

The final pattern is not a collection of randomly positioned dense domains. The radial density profile $\varphi$(r), plotted against the radial distance normalized by the lattice constant a, exhibits pronounced oscillations [Fig. 1(e)]. The first major peak occurs at r/a = 1, corresponding to the nearest-neighbor cell-center distance, while a secondary feature appears near r/a = $\sqrt{2}$, corresponding to the diagonal next-nearest-neighbor distance of a square lattice. A complementary fourfold bond-orientational analysis gives $\Psi_4 = 0.91$, and the two principal bond-direction families are separated by approximately 89.3° (Fig. S1), providing direct quantitative support for the fourfold symmetry. The lattice spacing is externally tunable: the lattice constant a decreases monotonically as the driving frequency is raised from 10 Hz to 22 Hz [Fig. 1(f)]. The dynamical origin of this frequency dependence is discussed below.

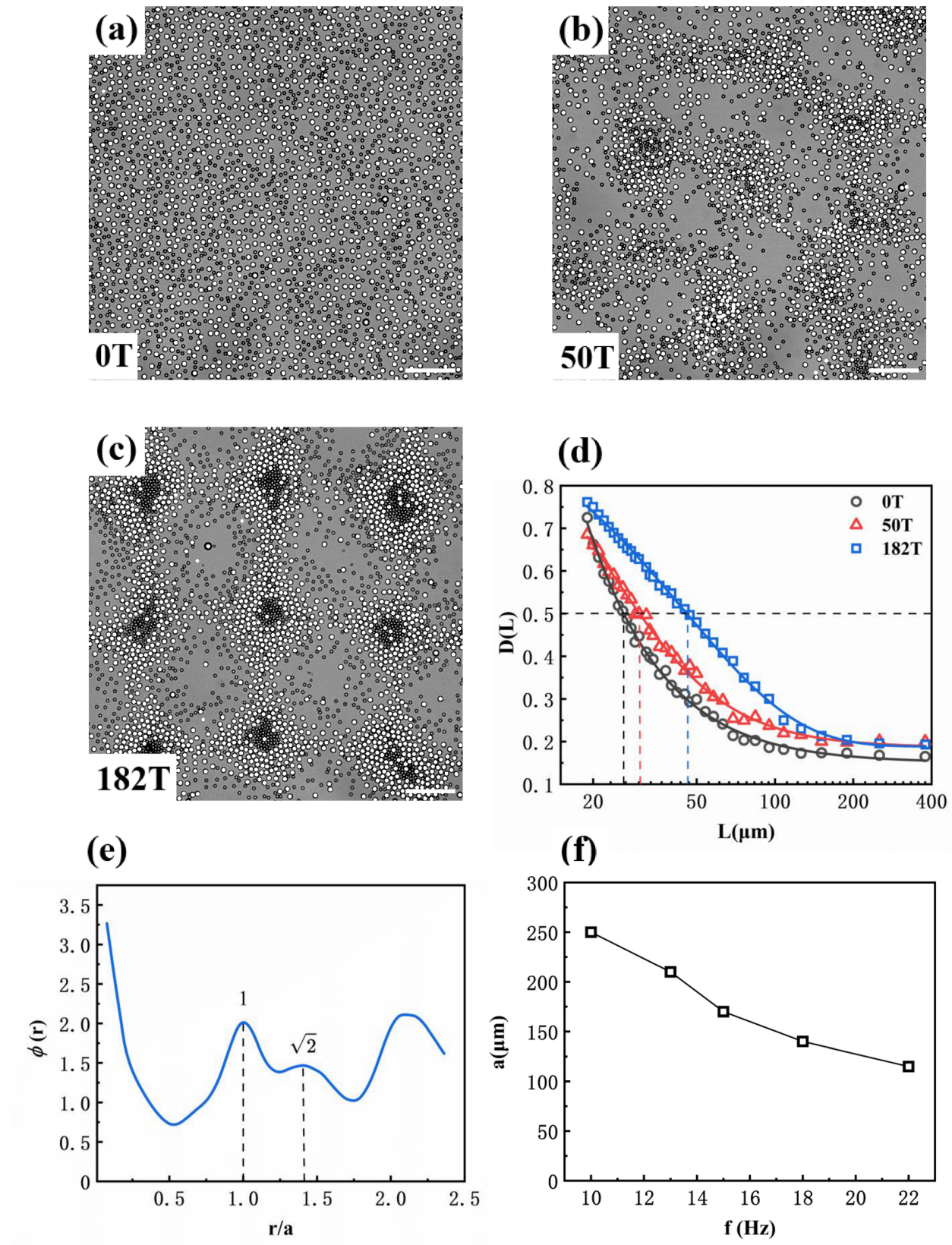

FIG. 1. Formation and structural characterization of the spatially demixed state under periodic square-wave electric-field driving ($E_p/E_c = 2$, f = 20 Hz, $\varphi = 0.175$ unless otherwise stated). (a)-(c) Particle distributions at 0T, 50T, and 182T, respectively. The 10 $\mu$m particles are bright/fluorescent and the 6 $\mu$m particles are dark. Scale bar: 100 $\mu$m. (d) Scale-dependent segregation parameter D(L) at different evolution times, shown with a linear vertical axis. The horizontal dashed line marks D(L) = 0.5, and the vertical dashed lines indicate the corresponding characteristic segregation scales. (e) Radial density profile $\varphi$(r) as a function of the dimensionless radial distance r/a. The marked positions r/a = 1 and r/a = $\sqrt{2}$ correspond to the nearest-neighbor and diagonal next-nearest-neighbor distances of a square lattice, respectively. (f) Lattice constant a as a function of driving frequency f at $E_p/E_c = 2$.

The geometrical origin of the fourfold organization can be further understood from the periodic redistribution of the dense regions. During the active stage, parti-cles are expelled from locally dense regions toward the surrounding dilute regions, whereas the density configuration is largely arrested during the inactive stage. A Voronoi tessellation constructed from the instantaneous high-density cluster centers shows that the cluster centers formed in the subsequent cycle are preferentially located near the Voronoi vertices of the preceding configuration (Fig. S5), i.e., near approximately equidistant void regions between neighboring clusters. The pattern therefore undergoes an alternating reconstruction between cluster centers and void centers, consistent with the dual-transformation framework reported for periodically driven Quincke colloids[38]. Since an ideal square lattice is self-dual, such a center-to-void transformation preserves fourfold symmetry while shifting the lattice spatially. This geometrical compatibility provides a natural explanation for the emergence and persistence of the dynamic square-lattice structure under periodic driving.

The ordered state remains dynamically active under the periodically switched electric field. Figure 2(a) shows the mean-squared displacements (MSDs) of the two particle species. At short times, both curves approximately follow the $t^2$ guide, indicating predominantly directed motion during an individual activation stage. At longer times, the overall growth approaches the t guide, indicating a crossover toward diffusive-like dynamics on the measured time scale. Superimposed on this growth are pronounced oscillations spanning successive driving periods, reflecting the periodic back-and-forth response to the square-wave driving. Because the corresponding displacement scale remains smaller than the lattice spacing, we use the MSD here to characterize the local periodic motion rather than as direct evidence for migration between lattice cells.

The trajectory maps over six driving periods directly visualize this periodic back-and-forth motion [Figs. 2(b)-2(d)]. All three panels correspond to exactly the same spatial region and the same 300-ms time interval. Figure 2(b) shows the combined trajectories of both particle species and is obtained by superposing the species-resolved trajectories shown in Figs. 2(c) and 2(d). Particle excursions are organized by the square-lattice pattern rather than distributed uniformly over the field of view. The trajectories of the 6 $\mu$m particles are concentrated mainly inside the lattice cells [Fig. 2(c)], whereas the trajectories of the 10 $\mu$m particles are preferentially distributed around the cell peripheries and grid-like boundaries [Fig. 2(d)]. This species-dependent spatial constraint is a direct dynamical signature of the compositionally segregated lattice.

To determine how the initially mixed population develops the species-dependent spatial organization, Fig. 3 compares the mean speed and the probability distribution P(S) of the single-cycle path length at three stages of pattern formation. At the early stage (2T) [Figs. 3(a) and 3(d)], both species respond periodically to the electric-field pulses, while the path-length distribution of the 6 $\mu$m particles is shifted to slightly larger S than that of the 10 $\mu$m particles. The smaller particles therefore display a modestly larger effective migration distance per cycle at this stage. The microscopic origin of this initial activity difference cannot be isolated from the present many-particle measurements. It likely reflects a combination of size-dependent Quincke rolling and near-electrode conversion, electrostatic and hydrodynamic interactions, particle-electrode coupling, collisions, and crowding effects. We therefore regard it as an experimentally observed effective activity contrast rather than attributing it to a single microscopic mechanism.

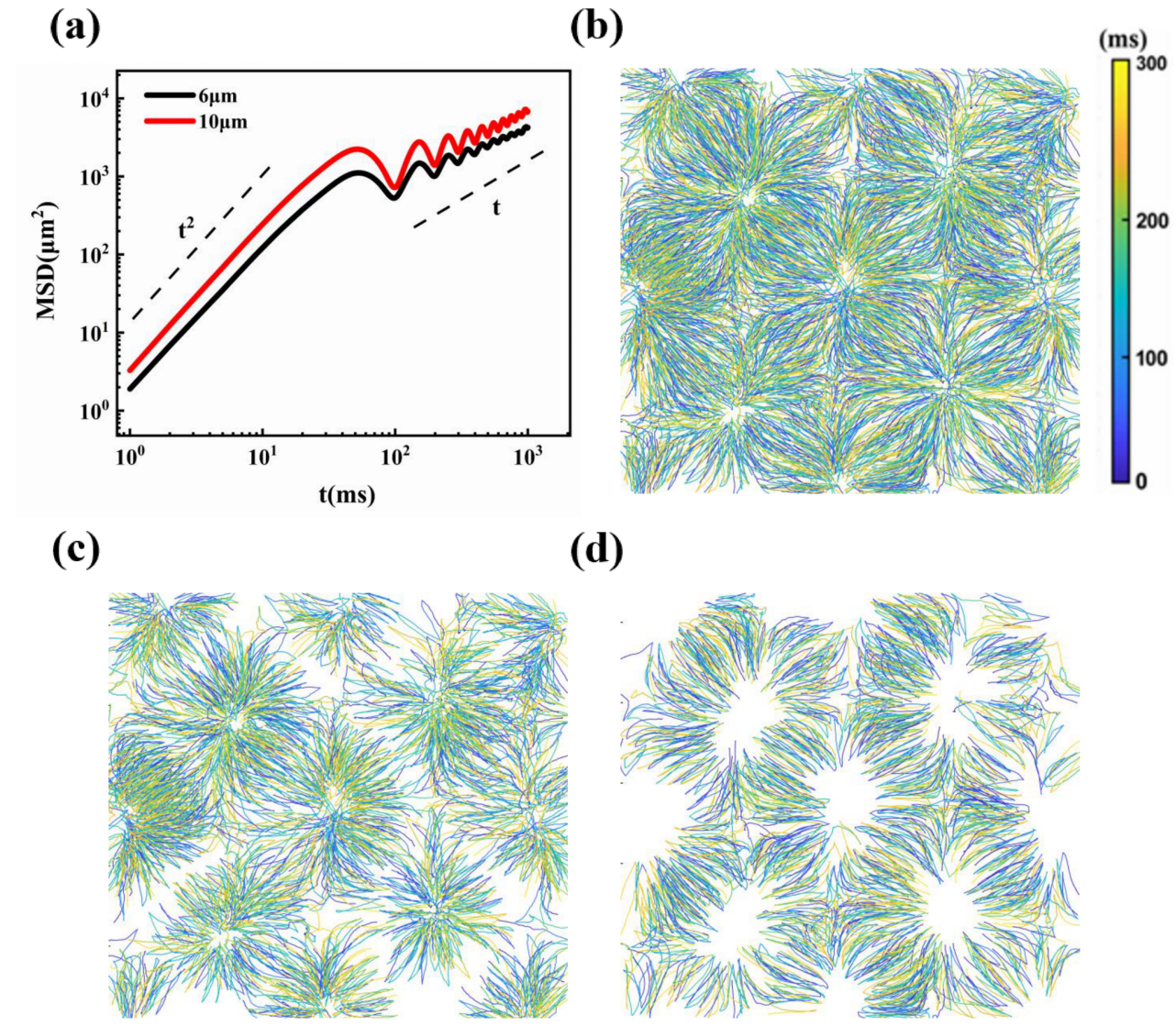

FIG. 2. Periodic back-and-forth motion of the particles ($E_p/E_c$ = 2, f = 20 Hz, $\varphi$ = 0.175). (a) Mean-squared displacements (MSDs) of the 6 $\mu$m and 10 $\mu$m particles. Dashed guides indicate $t^2$ and t scaling. (b) Combined particle trajectories over six driving periods (300 ms); color denotes elapsed time. (c) Species-resolved trajectories of the 6 $\mu$m particles. (d) Species-resolved trajectories of the 10 $\mu$m particles. Panels (b)-(d) correspond to exactly the same spatial region and time interval, and panel (b) is obtained by superposing the trajectories shown in panels (c) and (d).

At the intermediate stage (54T) [Figs. 3(b) and 3(e)], the two P(S) distributions approach each other as the compositional pattern becomes established. The evolving local environment therefore modifies the initially size-dependent dynamical response. At the late stage (130T) [Figs. 3(c) and 3(f)], the ordering reverses: the 10 $\mu$m particles display higher velocity plateaus and a path-length distribution shifted to substantially larger S. This reversal is consistent with the trajectory maps in Fig. 2. Once small-particle-rich cores and large-particle-rich peripheries have formed, the larger particles have more accessible space along the lattice boundaries, whereas the smaller particles become geometrically constrained inside the cells.

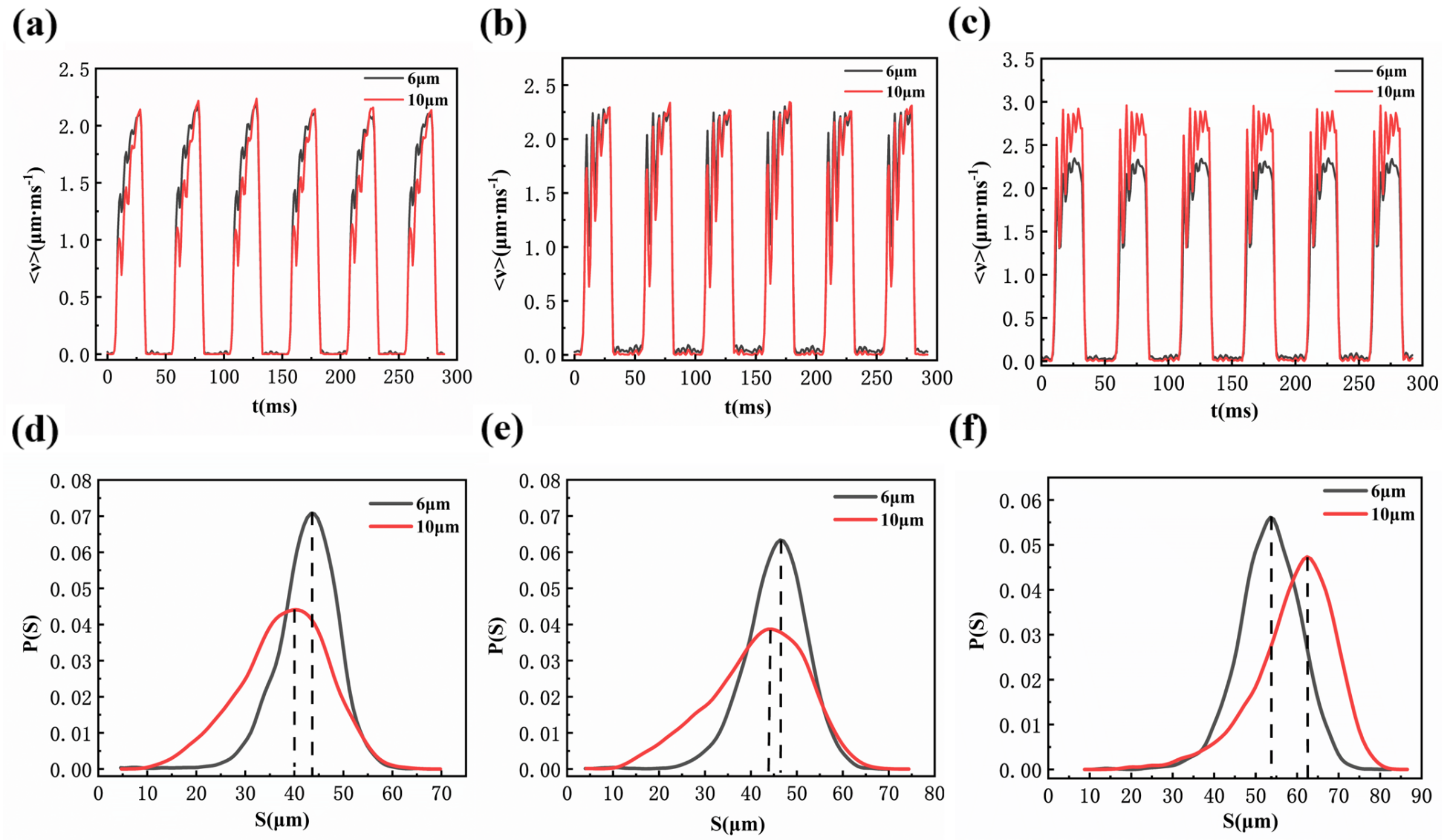

FIG. 3. Evolution of particle dynamics during formation of the spatially demixed structure ($E_p/E_c = 2$, f = 20 Hz, $\varphi = 0.175$). (a)-(c) Mean speeds of the 6 $\mu$m and 10 $\mu$m particles as functions of time at the early (2T), intermediate (54T), and late (130T) stages, respectively. (d)-(f) Corresponding probability distributions P(S) of the single-cycle path length S. Dashed vertical lines indicate the peak positions of the corresponding P(S) distributions.

The segregation process is therefore accompanied by a correlation between size-dependent migration and the structure generated during redistribution. Early in the evolution, the larger single-cycle displacement of the 6 $\mu$m particles is associated with the development of small-particle-rich cores. As segregation develops, 10 $\mu$m particles are displaced toward the periphery, where their motion becomes less constrained and the dynamical hierarchy reverses. Although the present measurements do not directly establish an escape-suppression mechanism, this late-time dynamical asymmetry is consistent with, and may contribute to, maintenance of the small-inside/large-outside composition. In this process the integrated path length S provides a more direct measure of spatial redistribution than the instantaneous speed alone.

Consistent with the frequency dependence of the lattice spacing shown in Fig. 1(f), the single-cycle migration lengths of both particle species decrease systematically with increasing driving frequency (Fig. 4). Because the square-wave field contains equal on and off halves, increasing f shortens the active interval available for migration within each cycle, thereby reducing the accessible displacement from a high-density region toward a neighboring low-density region. The direct a-$S_{eff}$ correlation and the corresponding dimensionless ratios are provided in Figs. S2 and S4, respectively.

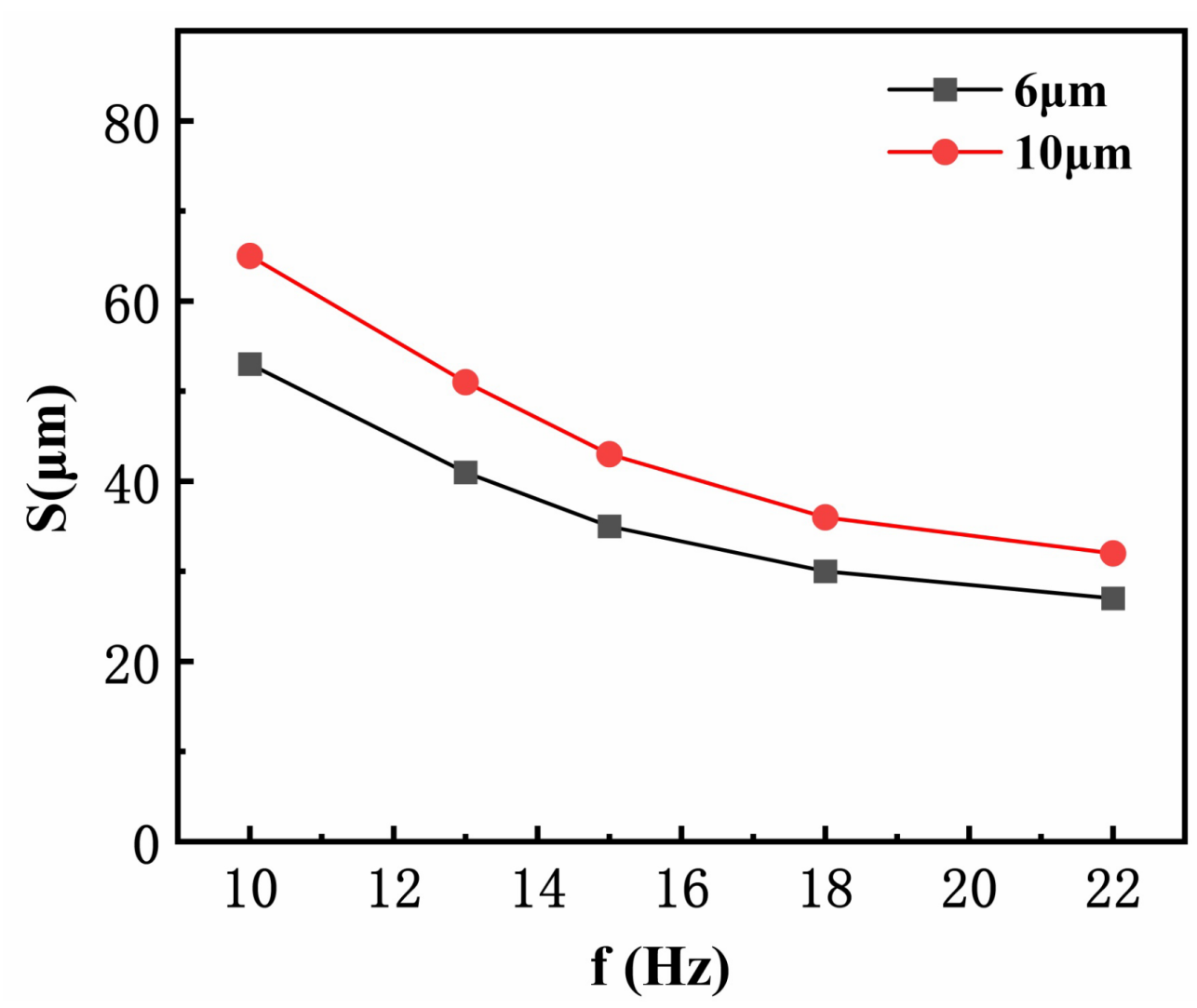

FIG. 4. Mean single-cycle path length S of the 6 $\mu$m and 10 $\mu$m particles as a function of the driving frequency f ($E_p/E_c$ = 2, $\varphi$ = 0.175). The path length decreases monotonically with increasing frequency for both particle species.

To test this connection directly, we define an effective mean single-cycle migration length $S_{eff} = (S_6 + S_{10})/2$ and plot the lattice constant a against $S_{eff}$ (Fig. S2). Over the investigated frequency range, a shows a strong approximately linear correlation with $S_{eff}$ ($R^2 = 0.977$). We do not imply a universal scaling law; rather, this comparison demonstrates directly that the characteristic lattice scale tracks the measured migration scale. A dimensionless comparison further shows that $S_{eff}$/a varies only from 0.219 to 0.257, with an average value of 0.235 (Fig. S4), supporting a close correspondence between the single-cycle migration scale and the characteristic lattice spacing. The periodically interrupted propulsion limits the distance over which particles can redistribute during each activation stage, thereby providing a natural finite length scale and limiting unrestricted coarsening toward macroscopic binary separation. Periodic forcing therefore provides a convenient control parameter for programming the length scale of the finite-scale demixed lattice.

In summary, we have investigated the nonequilibrium self-organization of a binary Quincke suspension containing equal numbers of 6 $\mu$m and 10 $\mu$m particles under periodic square-wave electric-field driving. Starting from a randomly mixed state, repeated activation and relaxation generate finite-scale compositional demixing and a dynamic square lattice in which small particles are concentrated near lattice-cell centers and large particles preferentially occupy the periphery. Particle-resolved dynamics show that the two species do not retain a fixed motility hierarchy: small particles have the larger single-cycle migration distance at early times, whereas the dynamical advantage reverses after the large particles are displaced to the less constrained lattice periphery. This evolving dynamical asymmetry is consistent with, and may contribute to, maintenance of the observed compositional organization. Increasing the driving frequency shortens the single-cycle path lengths of both species and is accompanied by a reduction of the lattice constant, providing a direct route to tune the pattern scale through temporal driving. These results demonstrate how particle-size heterogeneity and periodic activity control can cooperate to create programmable, compositionally structured active lattices.

*Materials and Methods.—*

**Experimental setup.** The experiments use a binary suspension of polystyrene spheres with diameters 6 $\mu$m and 10 $\mu$m dispersed in a weakly conducting AOT/hexadecane solution. The suspension is confined between two indium-tin-oxide-coated glass slides, and the particles sediment onto the lower electrode to form a quasi-two-dimensional layer. Above the Quincke threshold field $E_c$, spontaneous electrohydrodynamic rotation is converted into rolling motion along the electrode surface.[24, 25] A function generator and a high-voltage amplifier apply a periodic square-wave electric field that switches between a peak field $E_p > E_c$ and $E_g = 0$. The on and off stages each occupy one half of the driving period T = 1/f. Particle motion is recorded by high-speed optical microscopy and analyzed from particle-resolved trajectories.

**Segregation parameter.** To quantify compositional separation, the field of view is divided into square boxes of side length L. If $N_s$ and $N_l$ denote the numbers of 6 $\mu$m and 10 $\mu$m particles within a box, respectively, the scale-dependent segregation parameter is defined as

$$D(L) = \left\langle \frac{|N_l - N_s|}{N_l + N_s} \right\rangle . \qquad (1)$$

The angle brackets denote an average over all non-empty boxes; boxes containing no particles ($N_l + N_s = 0$), if any, are excluded from the averaging. D(L) approaches zero for a locally well-mixed binary population and increases as one species dominates a box. D(L) = 0.5 is used as a convenient intermediate operational threshold for defining a characteristic segregation scale rather than as a unique physical boundary. Thresholds D = 0.4, 0.5, and 0.6 give the same increasing trend (Table S1).

**Structural and dynamical analysis.** The spatial lattice is characterized by the radial density profile measured from high-density cell centers and by the nearest-neighbor lattice constant a. For dimensionless structural analysis, the radial distance is normalized by the corresponding lattice constant, r/a. Particle dynamics are characterized by the mean-squared displacement (MSD), the mean speed $\langle v \rangle$, and the single-cycle path length S obtained by summing the particle displacement over one active stage. The probability distribution P(S) is calculated separately for the two particle species at different stages of pattern formation. To compare migration and structural scales, the single-cycle path lengths are additionally normalized by a, giving $S_6$/a, $S_{10}$/a, and $S_{eff}$/a, where $S_{eff} = (S_6 + S_{10})/2$.

*Competing interests.—*

All authors declare that they have no competing interests.

*Data availability.—*

The data supporting this article, including Movie 1, are included as part of the Supplementary Information.

*Acknowledgments.—*

K.C., W.D.T. and T.H.Z. acknowledge financial support from the National Natural Science Foundation of China (Grant Nos. 21774091 and 22673081 to K.C.; Grant No. 22573069 to W.D.T.; Grant No. 12574230 to T.H.Z.).

*Author contributions.—*

T.H.Z., K.C. and W.D.T. designed and supervised the research. C.Y.Z., M.Y.Z. and X.Y.Z. performed the research and contributed equally. C.Y.Z., M.Y.Z.

and X.Y.Z. analyzed the experimental data with K.C., W.D.T. and T.H.Z. C.Y.Z. prepared the manuscript, and K.C., W.D.T. and T.H.Z. revised the manuscript. All authors discussed the results and approved the final manuscript.

---

* These authors contribute equally to this work.
† tianwende@suda.edu.cn
‡ kangchen@suda.edu.cn
§ zhangtianhui@suda.edu.cn

[1] Whitesides G M and Grzybowski B 2002 *Science* **295** 2418
[2] Grzelczak M, Vermant J, Furst E M and Liz-Marzán L M 2010 *ACS Nano* **4** 3591
[3] Lv J, Gao X, Han B, Zhu Y, Hou K and Tang Z 2022 *Nat. Rev. Chem.* **6** 125
[4] Ramaswamy S 2010 *Annu. Rev. Condens. Matter Phys.* **1** 323
[5] Marchetti M C, Joanny J F, Ramaswamy S, Liverpool T B, Prost J, Rao M and Simha R A 2013 *Rev. Mod. Phys.* **85** 1143
[6] Bowick M J, Fakhri N, Marchetti M C and Ramaswamy S 2022 *Phys. Rev. X* **12** 010501
[7] Bechinger C, Di Leonardo R, Löwen H, Reichhardt C, Volpe G and Volpe G 2016 *Rev. Mod. Phys.* **88** 045006
[8] Gompper G, Winkler R G, Speck T *et al.* 2020 *J. Phys.: Condens. Matter* **32** 193001
[9] Palacci J, Sacanna S, Steinberg A P, Pine D J and Chaikin P M 2013 *Science* **339** 936
[10] Theurkauff I, Cottin-Bizonne C, Palacci J, Ybert C and Bocquet L 2012 *Phys. Rev. Lett.* **108** 268303
[11] Bishop K J M, Biswal S L and Bharti B 2023 *Annu. Rev. Chem. Biomol. Eng.* **14** 1
[12] Zöttl A and Stark H 2023 *Annu. Rev. Condens. Matter Phys.* **14** 109
[13] Wu H, Shen Y, Sivagurunathan S, Weber M S, Adam S A, Shin J H, Fredberg J J, Medalia O, Goldman R and Weitz D A 2022 *Proc. Natl. Acad. Sci. USA* **119** e2115217119
[14] McCandlish S R, Baskaran A and Hagan M F 2012 *Soft Matter* **8** 2527
[15] Stenhammar J, Wittkowski R, Marenduzzo D and Cates M E 2015 *Phys. Rev. Lett.* **114** 018301
[16] Agudo-Canalejo J and Golestanian R 2019 *Phys. Rev. Lett.* **123** 018101
[17] You Z, Baskaran A and Marchetti M C 2020 *Proc. Natl. Acad. Sci. USA* **117** 19767
[18] Fruchart M, Hanai R, Littlewood P B and Vitelli V 2021 *Nature* **592** 363
[19] Maity S and Morin A 2023 *Phys. Rev. Lett.* **131** 178304
[20] Xie Y, Zhou X Y, Ni Q Y, Tian W D, Chen K and Zhang T H 2025 *Chin. Phys. B* **34** 126401
[21] Cates M E and Tailleur J 2015 *Annu. Rev. Condens. Matter Phys.* **6** 219
[22] Geyer D, Martin D, Tailleur J and Bartolo D 2019 *Phys. Rev. X* **9** 031043
[23] Buttinoni I, Bialké J, Kümmel F, Löwen H, Bechinger C and Speck T 2013 *Phys. Rev. Lett.* **110** 238301
[24] Bricard A, Caussin J B, Desreumaux N, Dauchot O and Bartolo D 2013 *Nature* **503** 95
[25] Bricard A, Caussin J B, Das D, Savoie C, Chikkadi V, Shitara K, Chepizhko O, Peruani F, Saintillan D and Bartolo D 2015 *Nat. Commun.* **6** 7470
[26] Lu S Q, Zhang B Y, Zhang Z C, Shi Y and Zhang T H 2018 *Soft Matter* **14** 5092
[27] Das D and Saintillan D 2013 *Phys. Rev. E* **87** 043014
[28] Morin A, Desreumaux N, Caussin J B and Bartolo D 2017 *Nat. Phys.* **13** 63
[29] Chen Y, Wang L and Zhang T H 2023 *Soft Matter* **19** 512
[30] He K X, Ni Q Y, Zhou X Y, Tian W D, Chen K and Zhang T H 2025 *Chin. Phys. B* **34** 076401
[31] Liu L, Wu J, Song C, Chen K, Tian W D and Zhang T H 2026 *Acta Phys. Sin.* **75** doi:10.7498/aps.75.20260416 (in Chinese)
[32] Du H T, Zhou X Y, Ni Q Y, Chen K, Tian W D and Zhang T H 2025 *Acta Phys. Sin.* **74** 116401 (in Chinese)
[33] Yang Y, Zhang M F, Zhu L and Zhang T H 2023 *Chin. Phys. Lett.* **40** 126401
[34] Liu Z T, Shi Y, Zhao Y, Chaté H, Shi X Q and Zhang T H 2021 *Proc. Natl. Acad. Sci. USA* **118** e2104724118
[35] Zhang B, Snezhko A and Sokolov A 2022 *Phys. Rev. Lett.* **128** 018004
[36] Zhang B, Glatz A, Aranson I S and Snezhko A 2023 *Nat. Commun.* **14** 7050
[37] Zhang M F, Fan B Y, Zhang C Y, Chen K, Tian W D and Zhang T H 2025 *Soft Matter* **21** 927
[38] Yang Y, Zhang Z C, Qi F and Zhang T H 2022 arXiv:2204.07717 [cond-mat.soft] (preprint)

**Supplementary Information**

**Complex pattern in periodically driven binary Quincke colloids**

Chuan Yu Zhang, Mu Yue Zhao, Xiao Yi Zhou, Wen-De Tian, Kang Chen, and Tian Hui Zhang

## Figure S1. Quantitative fourfold structural order

Nearest-neighbor bond directions extracted from the final ordered state give a fourfold bond-orientational order parameter $\Psi_4 = 0.91$. The two principal bond-direction families are separated by approximately 89.3°, close to the 90° separation expected for square symmetry.

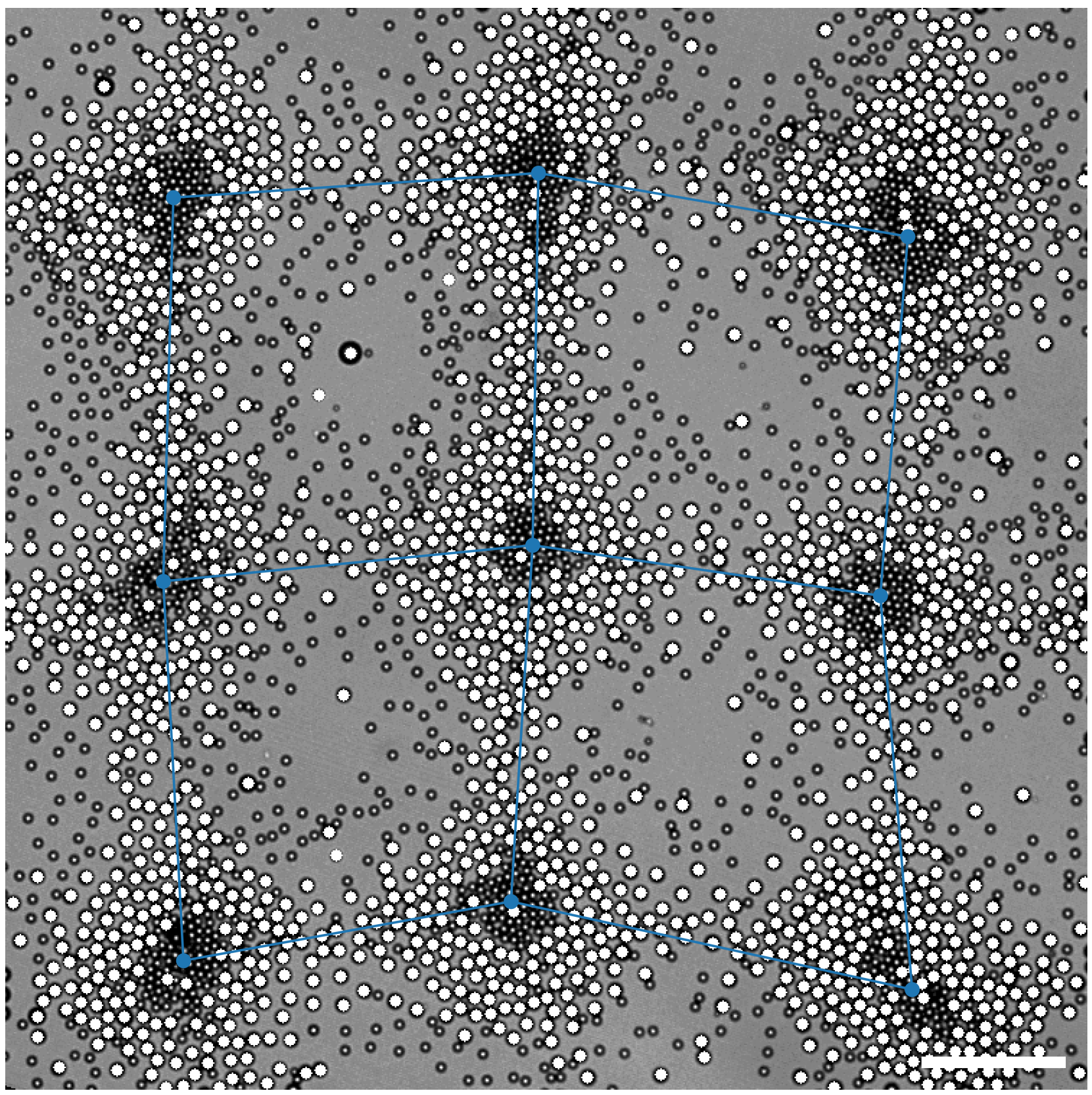

Fig. S1. Representative bond network used for the fourfold structural analysis of the final dynamic lattice. Blue points mark the selected high-density cell centers and blue segments indicate nearest-neighbor connections. The analysis gives $\Psi_4 = 0.91$, and the two principal bond-direction families are separated by approximately 89.3°.

## Table S1. Sensitivity of the characteristic segregation scale to the operational D(L) threshold.

| Threshold $D(L)$ | $0T$ ($\mu$m) | $50T$ ($\mu$m) | $182T$ ($\mu$m) |
|---|---|---|---|
| 0.4 | 32.8 | 41.2 | 65.5 |
| 0.5 | 26.3 | 28.9 | 46.5 |
| 0.6 | 21.8 | 22.9 | 32.5 |

The characteristic segregation scale increases from 0T to 182T for all three nearby thresholds, showing that the temporal trend does not depend on the precise choice D(L) = 0.5.

## Figure S2. Direct relation between lattice spacing and effective migration length

For the 1:1 binary mixture, the effective single-cycle migration length is defined as $S_{eff} = (S_6 + S_{10})/2$. Plotting a directly against $S_{eff}$ over the investigated frequencies gives a strong approximately linear correlation with $R^2 = 0.977$. This plot is used as a direct correlation analysis rather than as evidence for a universal scaling law.

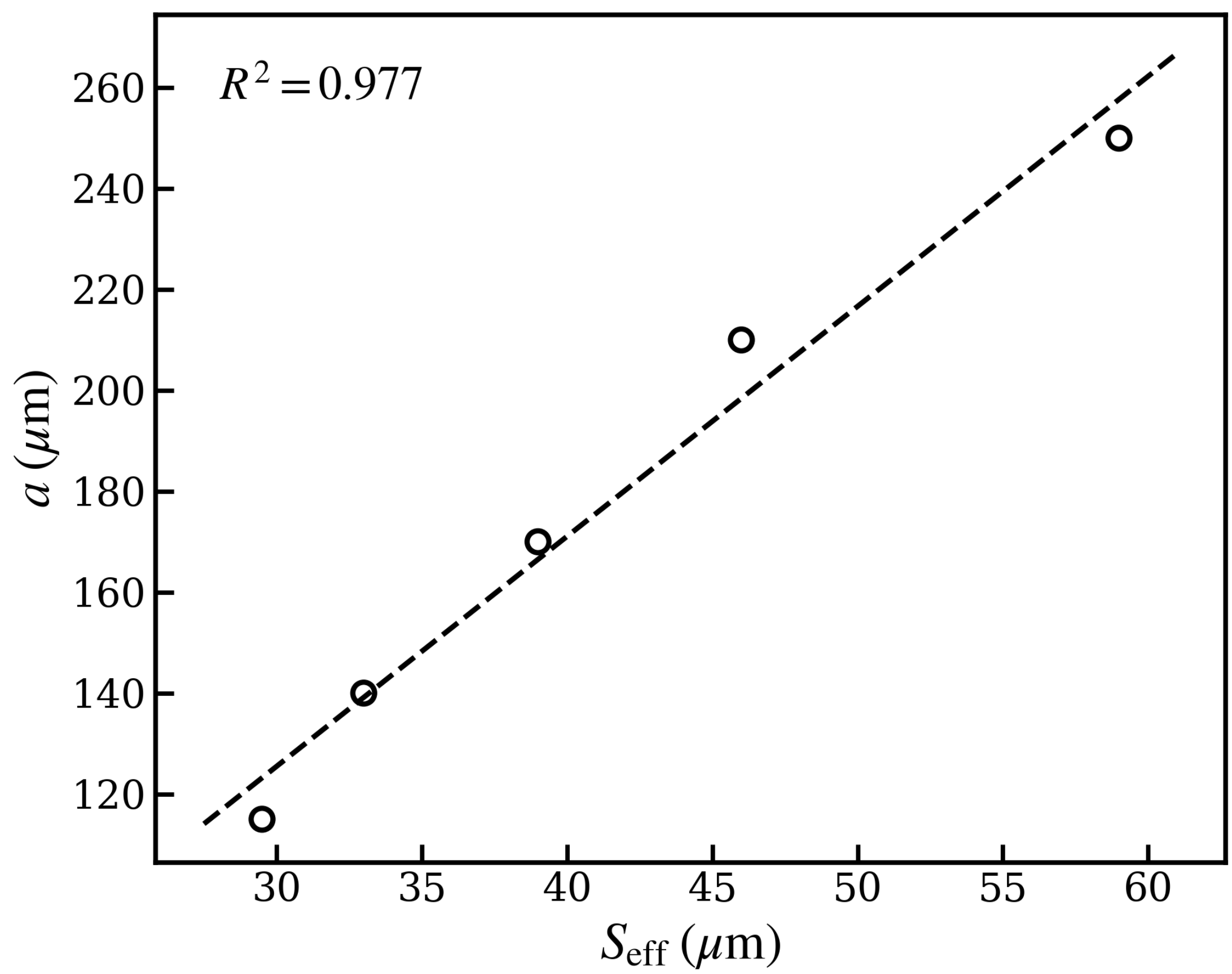

Fig. S2. Lattice constant a as a function of the effective single-cycle migration length $S_{eff} = (S_6 + S_{10})/2$. The dashed line is a linear fit ($R^2 = 0.977$).

## Figure S3. Species-resolved radial density profiles

To complement the total radial density profile in Fig. 1(e), the 6 $\mu$m and 10 $\mu$m particles were analyzed separately using the same set of high-density cell centers. The smaller particles show stronger enrichment near the cell centers, whereas the larger particles are relatively enriched farther from the centers, consistent with the small-inside/large-outside compositional organization. In Fig. S3, the radial coordinate is normalized by the corresponding lattice spacing, r/a.

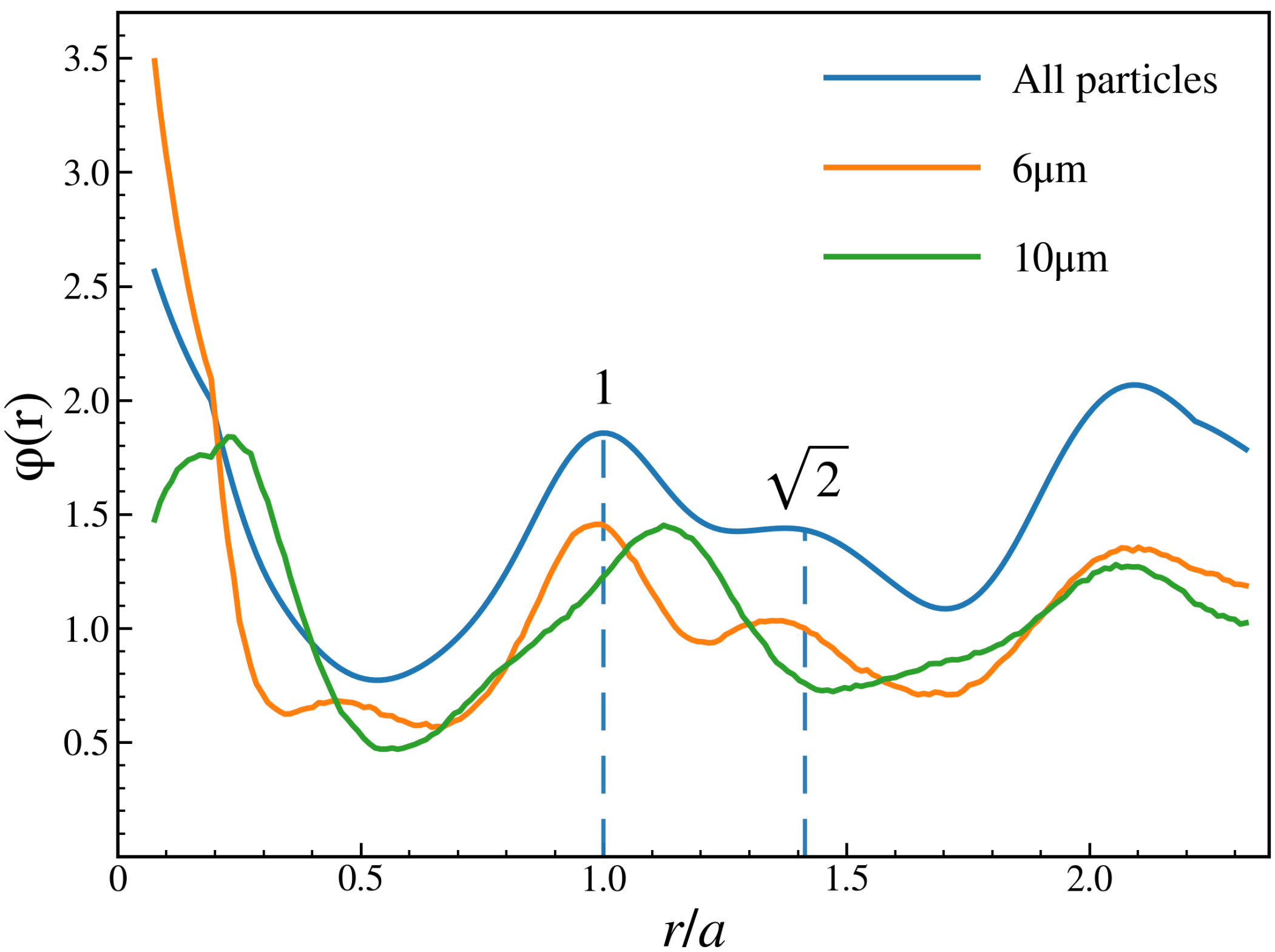

Fig. S3. Total and species-resolved normalized radial density profiles as functions of the dimensionless distance r/a. Dashed lines mark r/a = 1 and r/a = $\sqrt{2}$.

## Figure S4. Dimensionless comparison of migration and structural scales

The ratios $S_6$/a, $S_{10}$/a, and $S_{eff}$/a are compared across the investigated driving frequencies. $S_{eff}$/a varies from 0.219 to 0.257 with an average value of approximately 0.235, supporting a close proportional correspondence between the single-cycle migration scale and the characteristic lattice spacing.

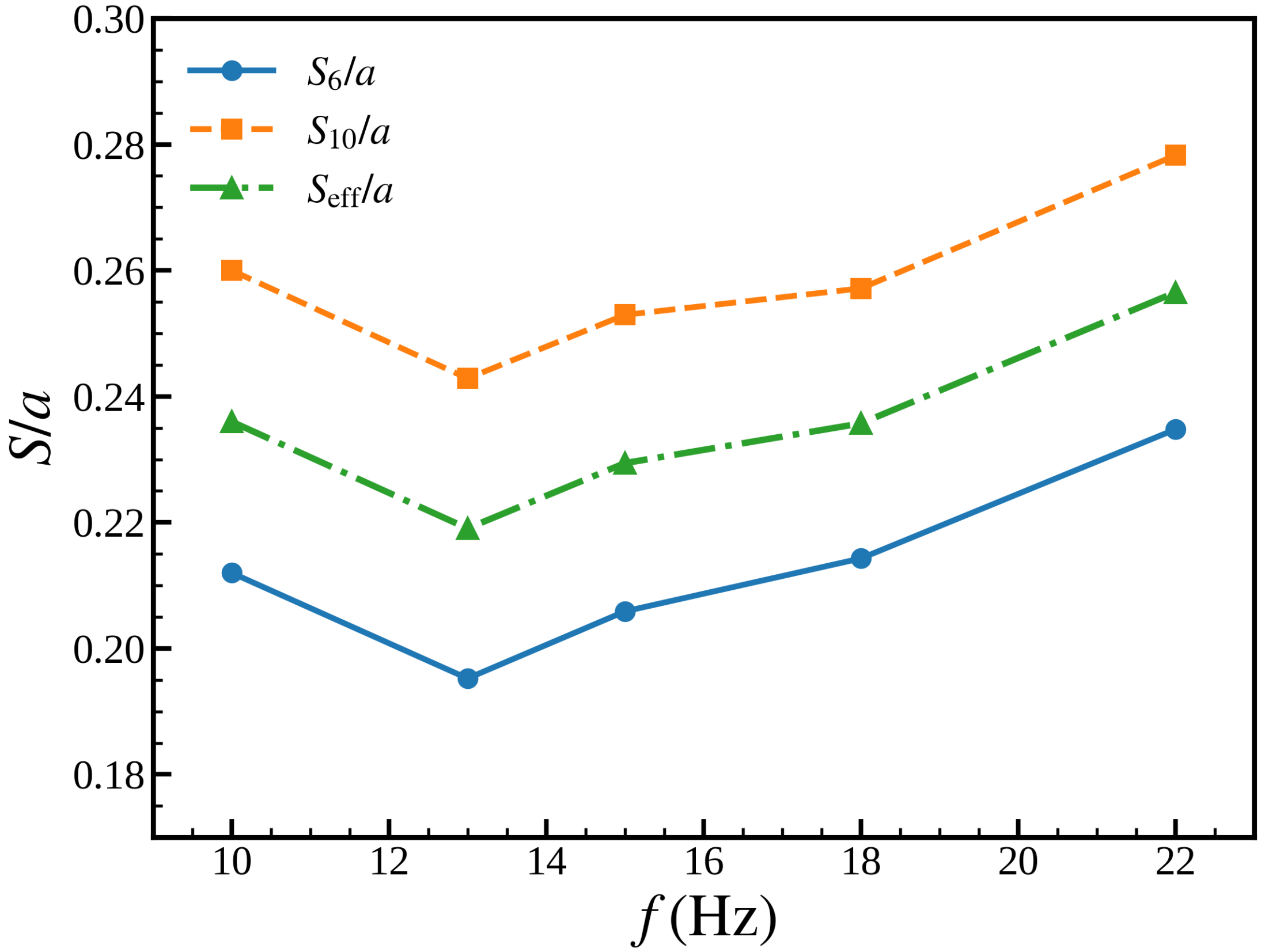

Fig. S4. Dimensionless ratios $S_6$/a, $S_{10}$/a, and $S_{eff}$/a as functions of the driving frequency f, where $S_{eff} = (S_6 + S_{10})/2$.

## Figure S5. Voronoi analysis of the dynamic square-lattice structure

The instantaneous high-density cluster centers are used as seeds for a Voronoi tessellation. In the subsequent cycle, newly formed cluster centers are preferentially located near the Voronoi vertices of the preceding configuration, indicating an alternating reconstruction between cluster centers and approximately equidistant void regions.

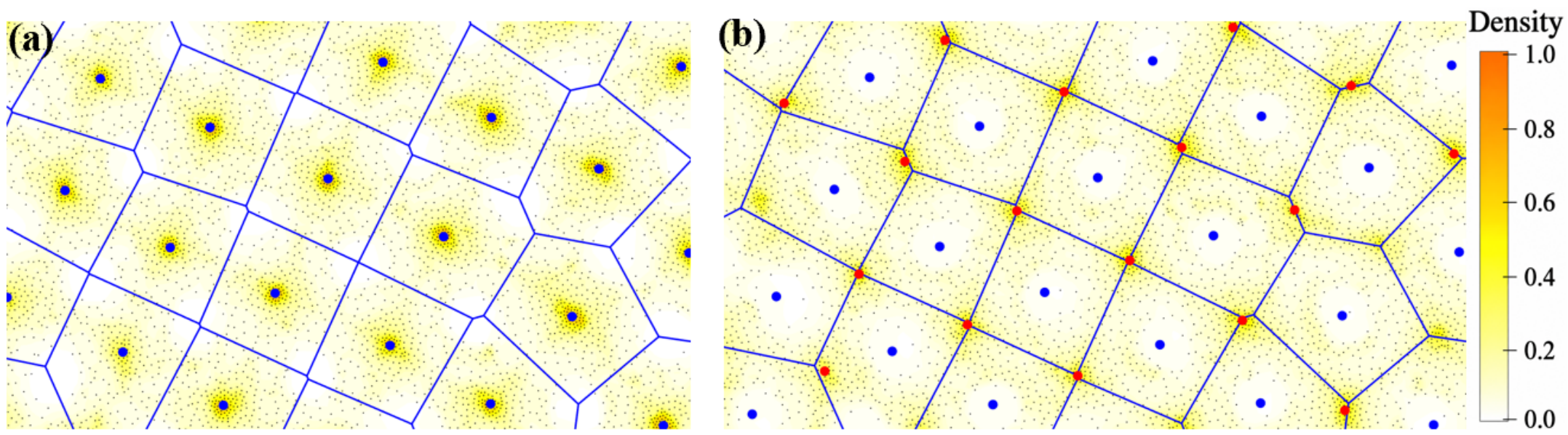

Fig. S5. Voronoi analysis of the dynamic square-lattice structure ($E_p/E_c = 2$, f = 10 Hz, $\varphi = 0.175$). (a) Voronoi tessellation constructed using the instantaneous centers of the high-density clusters (blue dots) as seeds. The small black dots indicate particle positions, and the background color represents the local density. (b) The centers of the high-density clusters in the current configuration (blue dots) and those formed in the subsequent cycle (red dots) are shown together with the same Voronoi tessellation. The newly formed cluster centers are preferentially located near the Voronoi vertices of the preceding configuration.

## Movie 1

Movie 1: Formation of the spatially demixed square lattice. $E_p/E_c = 2$, f = 20 Hz, and $\varphi = 0.175$. The movie shows the evolution of the binary Quincke suspension from the initially mixed state to the spatially demixed dynamic square lattice over 182 driving cycles (9.1 s of real experimental time). The movie is played back at 25 frames per second for approximately 29 s. The field of view is 842 $\mu$m $\times$ 842 $\mu$m. Scale bar: 100 $\mu$m.